\documentclass[sn-mathphys,Numbered]{sn-jnl}

\usepackage{graphicx}%
\usepackage{multirow}%
\usepackage{amsmath,amssymb,amsfonts}%
\usepackage{amsthm}%
\usepackage{mathrsfs}%
\usepackage[title]{appendix}%
\usepackage{xcolor}%
\usepackage{textcomp}%
\usepackage{manyfoot}%
\usepackage{booktabs}%
\usepackage{algorithm}%
\usepackage{algorithmicx}%
\usepackage{algpseudocode}%
\usepackage{listings}%
\usepackage{caption}
\usepackage{subcaption}
\usepackage{ulem}

\usepackage{changes}

\begin{document}

\title[Article Title]{Design and Simulation of TiN-Based Suspended Meander Kinetic Inductance Detectors (KIDs) for Visible
and Near-Infrared Astronomy Applications
}


\author*[1]{\fnm{Maria Appavou} \sur{}}\email{maria.appavou@obspm.fr}

\author[2]{\fnm{Lucas Ribeiro} \sur{}}\email{lucasrib@hotmail.fr}
\equalcont{These authors contributed equally to this work.}

\author[1]{\fnm{Paul Nicaise} \sur{}}\email{paul.nicaise@obspm.fr}
\equalcont{These authors contributed equally to this work.}

\author[1]{\fnm{Jie Hu} \sur{}}\email{jie.hu@obspm.fr}
\equalcont{These authors contributed equally to this work.}

\author[3]{\fnm{Jean-Marc Martin} \sur{}}\email{jeanmarc@caltech.edu}
\equalcont{These authors contributed equally to this work.}

\author[1]{\fnm{Josiane Firminy} \sur{}}\email{josiane.firminy@obspm.fr}
\equalcont{These authors contributed equally to this work.}

\author[1]{\fnm{Christine Chaumont} \sur{}}\email{christine.chaumont@obspm.fr}
\equalcont{These authors contributed equally to this work.}

\author[4]{\fnm{Piercarlo Bonifacio} \sur{}}\email{piercarlo.bonifacio@obspm.fr}
\equalcont{These authors contributed equally to this work.}

\author[1]{\fnm{Faouzi Boussaha} \sur{}}\email{faouzi.boussaha@obspm.fr}
\equalcont{These authors contributed equally to this work.}

\affil*[1]{\orgdiv{GEPI}, \orgname{Observatoire de Paris, Universit\'e PSL, CNRS}, \orgaddress{\street{77 Avenue Denfert-Rochereau}, \city{Paris}, \postcode{75014}, \country{France}}}

\affil[2]{\orgname{Université d'Evry Val d'Essonne}, \orgaddress{\street{23 Boulevard François Mitterand}, \city{Evry-Courcouronnes}, \postcode{91000}, \country{France}}}

\affil[3]{\orgname{California Institute of Technology}, \orgaddress{\street{1200 E. California Blvd.}, \city{Pasadena}, \postcode{91125}, \state{California}, \country{USA}}}

\affil*[4]{\orgdiv{GEPI}, \orgname{Observatoire de Paris, Universit\'e PSL, CNRS}, \orgaddress{\street{5 Place Jules Janssen}, \city{Meudon}, \postcode{92195}, \country{France}}}

\abstract{We report on simulations of a novel design of optical titanium nitride (TiN)-based Kinetic Inductance Detectors (KIDs) in order to improve their response to optical photons. We propose to separate the meander from the substrate to trap hot phonons generated by optical photons, preventing their rapid propagation through the substrate. These phonons would in turn contribute to the breaking of more Cooper pairs, thereby increasing the response of the detector. In our design, the meander is suspended a few hundred nanometers above the substrate. Furthermore, reflective gold (Au) or aluminum (Al)-based layers can be placed under the meander to improve photon coupling in the optical wavelengths.}

\keywords{KID, Suspended Meander, TiNx, Superconductor, Optical and Near-Infrared}



\maketitle

\section{Introduction}\label{sec1}

Kinetic Inductance Detectors (KIDs) through their low noise, high readout speed, and large-scale multiplexing are cutting-edge sensors and competitive with other cryogenic detectors such as Superconducting Tunnel Junctions (STJs) and Transition Edge Sensors (TESs) for submillimeter/THz to UV ranges \cite{Ulbricht_2021}. However, the optimal operation of optical KIDs in spectroscopy is currently hampered by low-measured spectral resolutions. These remain much lower than those predicted by theory given by $R \approx 1/2.355 \sqrt{(\eta h \nu/F \Delta)}$ \cite{DeVisser_2021, Zobrist_2022}, where $\eta$ is the conversion efficiency of quasiparticles, which is typically of the order of 0.57, $h \nu$ is the photon energy, $F$ is the Fano factor, generally  $\sim$ 0.2, and $\Delta \approx 1.76 k_B T_c$ is the energy gap of the superconductor. $T_c$ is its critical temperature and $k_B$ is the Boltzmann constant. The cause of this difference is still not well understood, but studies currently being carried out indicate that it would be due to various sources of noise, whether generated by the readout electronics or by two-level system (TLS) noise featured on the interface between the dielectrics and the superconductor \cite{Gao_2008}, which also reduces the response to optical photons \cite{Hu2021, Boussaha_2023}.

As $R$ is signal-to-noise ratio-dependent \cite{Boussaha_2023}, in order to improve the energy-resolving power, one can either decrease the noise or increase the signal response, or both. When it is difficult to further mitigate TLS noise effects, the signal-to-noise ratio, and thus $R$, can be improved by increasing the signal response. Photon absorption induces a down-conversion cascade involving electron-phonon interaction and hot phonons, which have a higher energy than the Debye energy, will rapidly escape from the superconducting thin film to the substrate and significantly limit the energy-resolving power \cite{Kozorezov_2007}. A few solutions were explored by other research groups such as the addition of a superconducting thin layer featuring a lower Debye energy \cite{Zobrist_2022}, between the meander and the substrate, or suspending the meander on a thin membrane \cite{DeVisser_2021} in order both to trap phonons within the meander to enhance the Cooper pairs breaking. In this study, we propose to go one step further with the membrane solution and fully suspend the meander, without using a membrane, to further enhance the trapping of hot phonons and increase the response. This design would also allow us to place a reflector under the suspended meander to improve the optical coupling and increase the quantum efficiency (QE). Here, we assume that photons that are not absorbed by the  TiN-based meander can pass through before being reflected by the reflector towards the latter.

\raggedbottom

\section{Design}\label{sec2}

\begin{figure}
\centering
\includegraphics[width=0.9\textwidth]{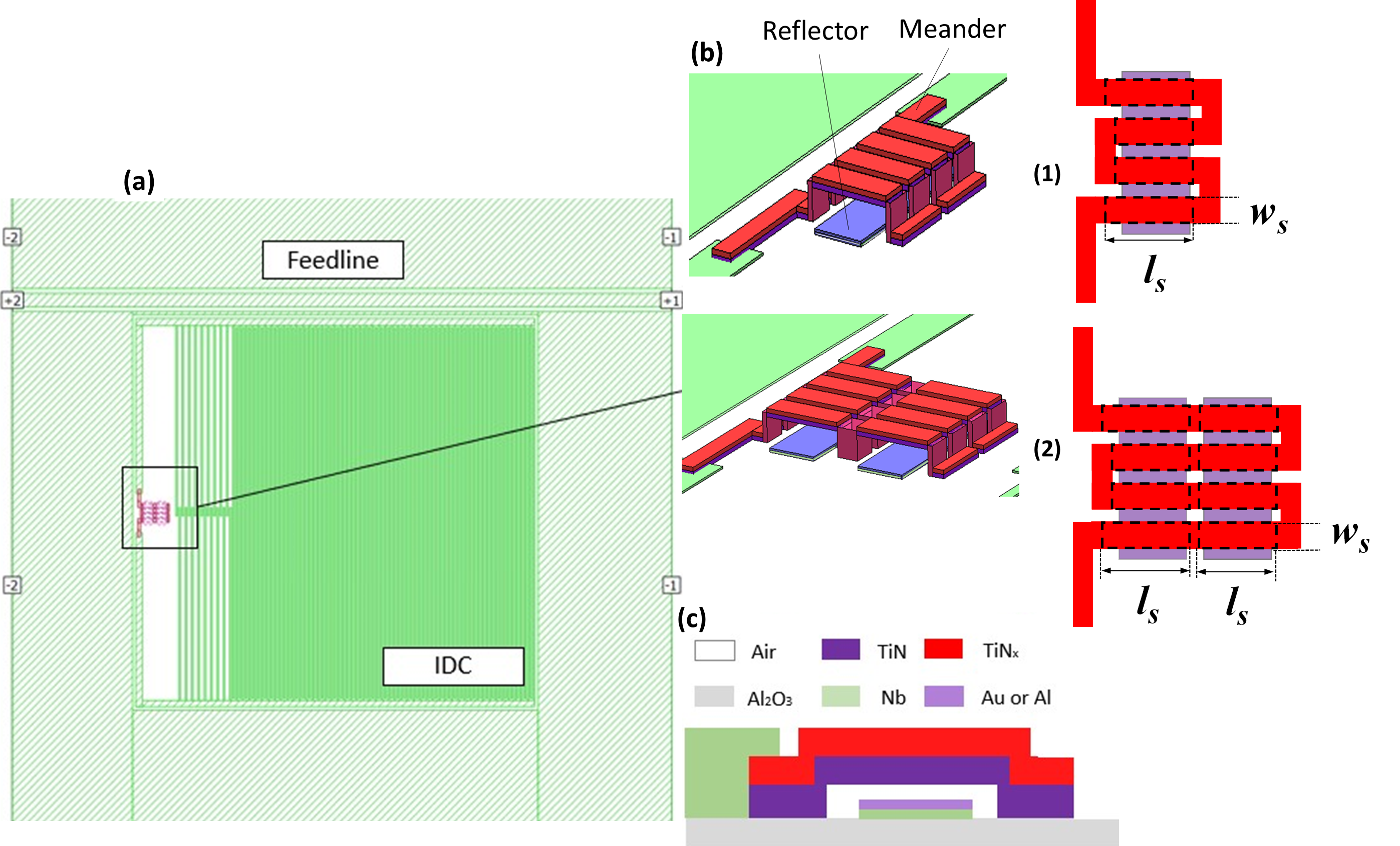}
\caption{(a) Top view of a suspended meander KID with a $300\times300~\mu\text{m}^2$ interdigitated capacitor (IDC) and a $22\times14~\mu \text{m} ^2$ meandered inductor. (b) Perspective (right) and top (left) views of the meander along with a reflector (1). For meanders with long strips, they can be divided into shorter sections (two here) which are easier to suspend (2). The dashed lines on top views indicate the suspended areas. (c) Side view of the suspended meander with reflector and the associated materials. For clarity, thicknesses are not to scale.}\label{fig1}
\end{figure}

Using SONNET software \cite{Sonnet}, our simulations are carried out with a sapphire ($\text{Al}_2\text{O}_3$) substrate whose relative dielectric constant $\epsilon_{r, Al_2O_3}$ is fixed at 9.9. Due to its large 10 eV energy gap, the sapphire is transparent at optical wavelengths, which means that, compared to silicon, the response of the substrate at these wavelengths is considerably reduced.
As illustrated in Fig.~\ref{fig1}, one pixel consists of an interdigitated capacitor (IDC) with a dimension of $300\times300~\mu \text{m}^2$ with 1~$\mu$m wide fingers and 1~$\mu$m gaps. 

Almost the entire meander can be suspended above a single rectangular reflector (Fig.~\ref{fig1}-b1), where only the edges are maintained in contact with the substrate. These contact points act as anchor points. The reflector can also be divided into smaller rectangles (Fig.~\ref{fig1}-b2) which allows shorter sections to be suspended if it proves difficult to do so with long strips. In this case, the meander will also be in contact with the substrate between the reflectors. It is obvious that contact areas must be as small as possible.

In this study, we consider the configuration with the split reflector (Fig. \ref{fig1}-b2). The meandered inductor has a total dimension of $l_m \times w_m~=~22\times14~\mu\text{m}$ and 1~$\mu$m gaps. The resonator is capacitively coupled to a 50~$\Omega$ coplanar waveguide (CPW) feedline with a 10~$\mu$m width conducting line separated by 3.5~$\mu$m width from both ground planes. 

\begin{table}[h]
\caption{Thickness, kinetic inductance, critical temperature, electrical resistivity of the different layers of the absorber and reflector.}\label{tab1}
\begin{tabular*}{\textwidth}{@{\extracolsep\fill}lcccccc}
\toprule%
Layer (Role) & $t$ [nm] & $L_k$ [pH/$\square$] & $T_c$ [K] & $\rho_{elec}$ [$\mu\Omega\cdot \text{cm}$] \\
\midrule
Nb \tnote{5}  & 40 & 0.2 & 9.2 & 15 \\
Au \cite{Matula_1979} or Al \cite{Desai_James_Ho_1984} (Reflector) \tnote{6}  & 30 & -  & - & 0.022 or 0.0001 \\ 
Vacuum & 100-250 &  - &  - & - \\
TiN (Absorber)\tnote{5} & 20 & 36 & 4.6  & 270 \\
$\text{TiN}_x$ (Absorber)\tnote{5} & 50-150 & 22-66 & 0.85 & 231 \\
\botrule
\end{tabular*}
\begin{tablenotes}
    \item[5] On SONNET, the kinetic inductance is defined as $L_{k} = \hbar\rho_n/\pi\Delta\text{t}$.
    \item[6] Simulated as normal metals with electrical resistivity $\rho_{elec}$ at $T\approx$ 1 K.
\end{tablenotes}\label{table1}
\end{table}

\begin{figure}[h]
\begin{subfigure}{0.5\textwidth}
\hspace{-0.3cm}
\includegraphics[width=0.8\linewidth, height=5cm]{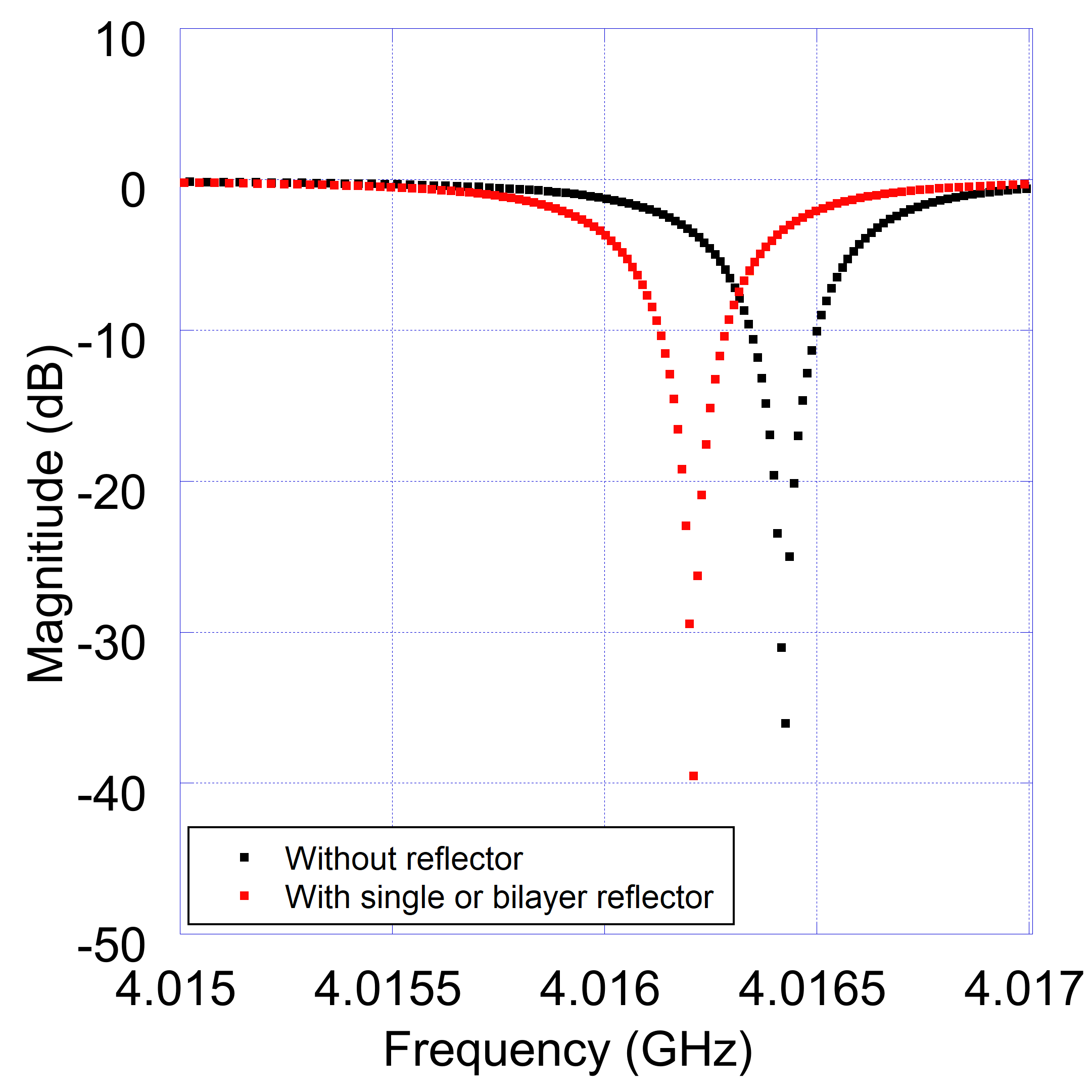} 
\caption{}
\label{fig2:subim1}
\end{subfigure}
\begin{subfigure}{0.8\textwidth}
\hspace{-1cm}
\includegraphics[width=0.7\linewidth, height=6cm]{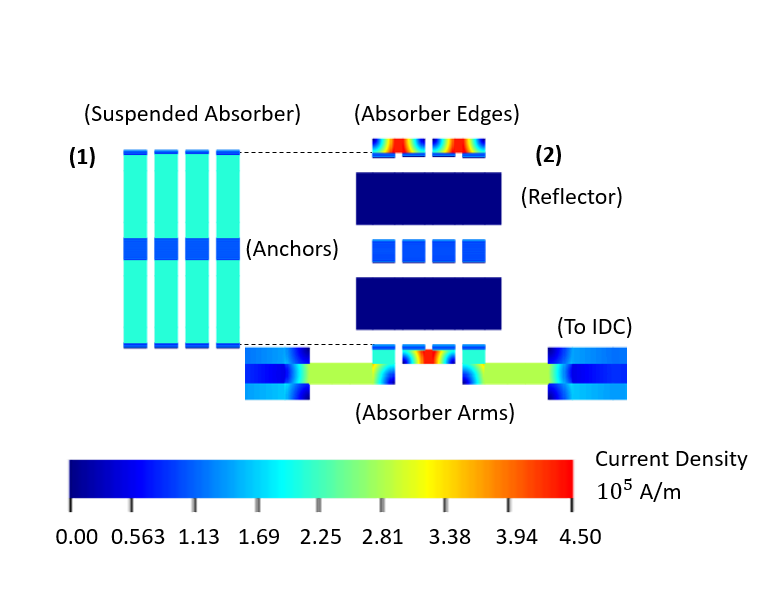} 
\caption{}
\label{fig2:subim2}
\end{subfigure}
\caption{(a) Frequency-domain representation of the transmission parameter $S_{21}$ for different configurations with vacuum gap (single and bilayer reflector configurations responses overlap). (b) Top view of the meander part showing the distribution of the current density in the suspended meander. (1) shows the suspended meander strips with a current density above $10^5$ A/m and (2) close to 0 A/m in the reflector.}
\label{fig:image2}
\end{figure}

The meander consists of a sub-stoichiometric $\sim$1~K TiN$_x$ thin film and both the IDC and feedline are made of a single niobium (Nb) thin film. Nb has a lower surface impedance making it easier to build the 50 $\Omega$ feedline. On the other hand, its higher energy gap than that of TiN prevents the diffusion of quasiparticles out of the TiN-based meander where they are generated. The diffusion of quasiparticles out of the absorber can induce pulse height variations \cite{DeVisser_2021}. A thin layer of stoichiometric $\sim$4~K TiN is added under the  TiN$_x$ layer to reduce the TLS noise by preventing oxidation when suspended. Furthermore, as the measured $T_c$ of the $\text{TiN}_x$ layer which meets the optimal mechanical stress parameters to achieve suspended superconducting structures \cite{Boussaha_2020} is around 0.8~K, the addition of the $\text{TiN}$ allows the overall increase of the meander $T_c$ up to $\sim$ 2 K. In fact, no resonance or very shallow resonances ($<2$~dB) were observed with our KIDs made from TiN$_x$ with a low $T_c~(<1$~K), probably due to high resistive losses. According to our first fabrication process optimizations to assess the best suspension conditions, we found that the meanders can easily collapse for strips of thickness less than  100 nm and length longer than 12 $\mu\text{m}$. In this case, the length and the thickness of the meander strip to be suspended are set at $l_s= 8~\mu\text{m}$ and $t_{meander}=120~\text{nm}$, respectively. The latter consists of 100 nm of $\text{TiN}_x$ and 20 nm of $\text{TiN}$, corresponding to respectively $L_{k,TiN_x}=33$ pH/$\square$ and $L_{k,TiN}=36$ pH/$\square$. Furthermore, the main parameters of the meander are stated in Tab.~\ref{tab1} as a baseline of the preliminary simulations. Also according to experimental investigations, the addition of a thin 20~nm-thick TiN layer does not alter the stress parameters of the $\text{TiN}_x$ layer required to be suspended. The strip width is set at $w_s = 2.7~\mu\text{m}$. Each strip is thus suspended over $2\times8~\mu\text{m}$-length. The contact area between the superconductor and the substrate is limited to $1.75~\times~2.7~\mu\text{m}^2$ per strip at edges and $2.25 \times 2.7~\mu\text{m}^2$ per strip at the center. Our fabrication process will be published somewhere else. 

\begin{figure}[h]
\begin{subfigure}{0.5\textwidth}
\includegraphics[width=0.9\linewidth, height=6cm]{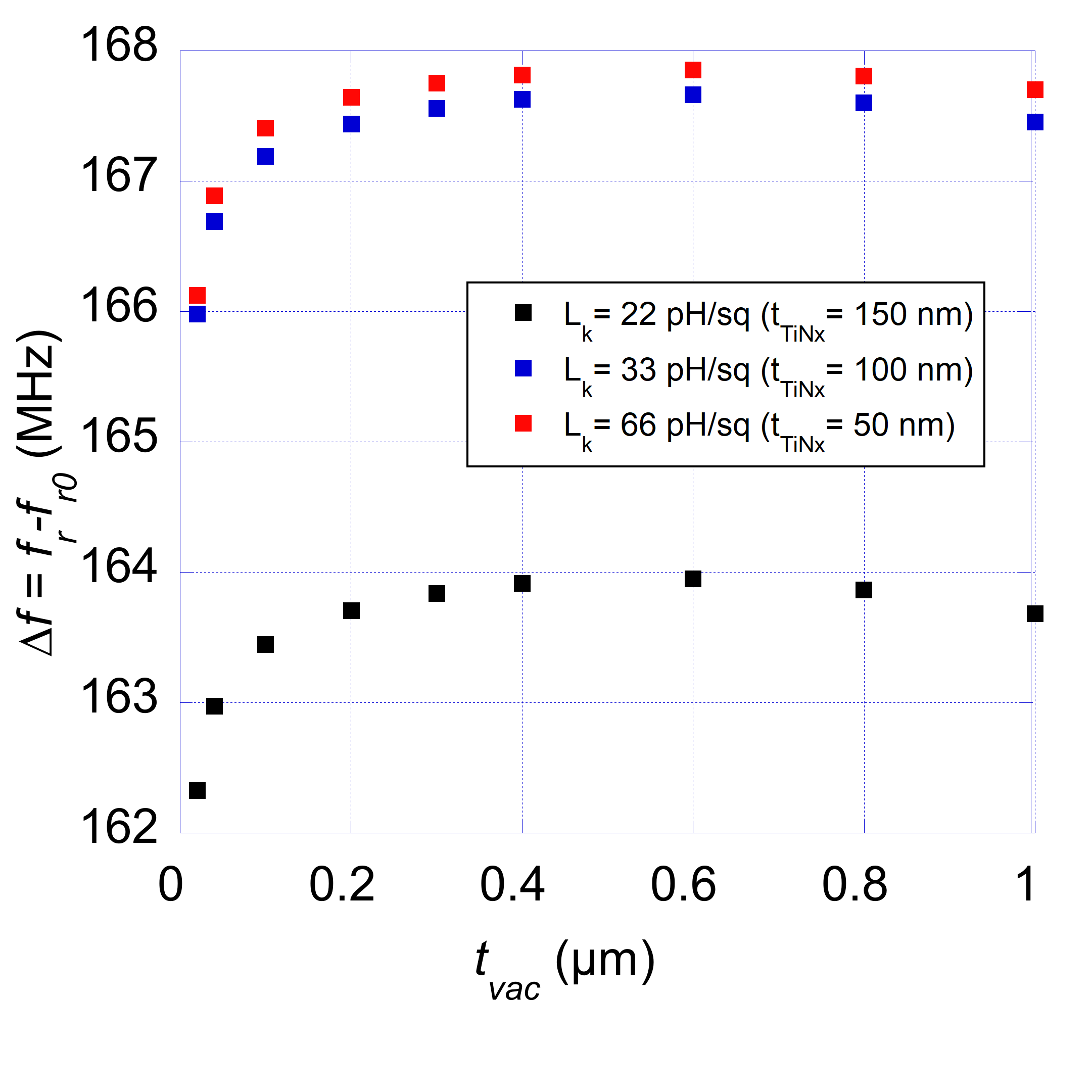} 
\caption{}
\label{fig3:subim1}
\end{subfigure}
\begin{subfigure}{0.5\textwidth}
\includegraphics[width=0.9\linewidth, height=6.1cm]{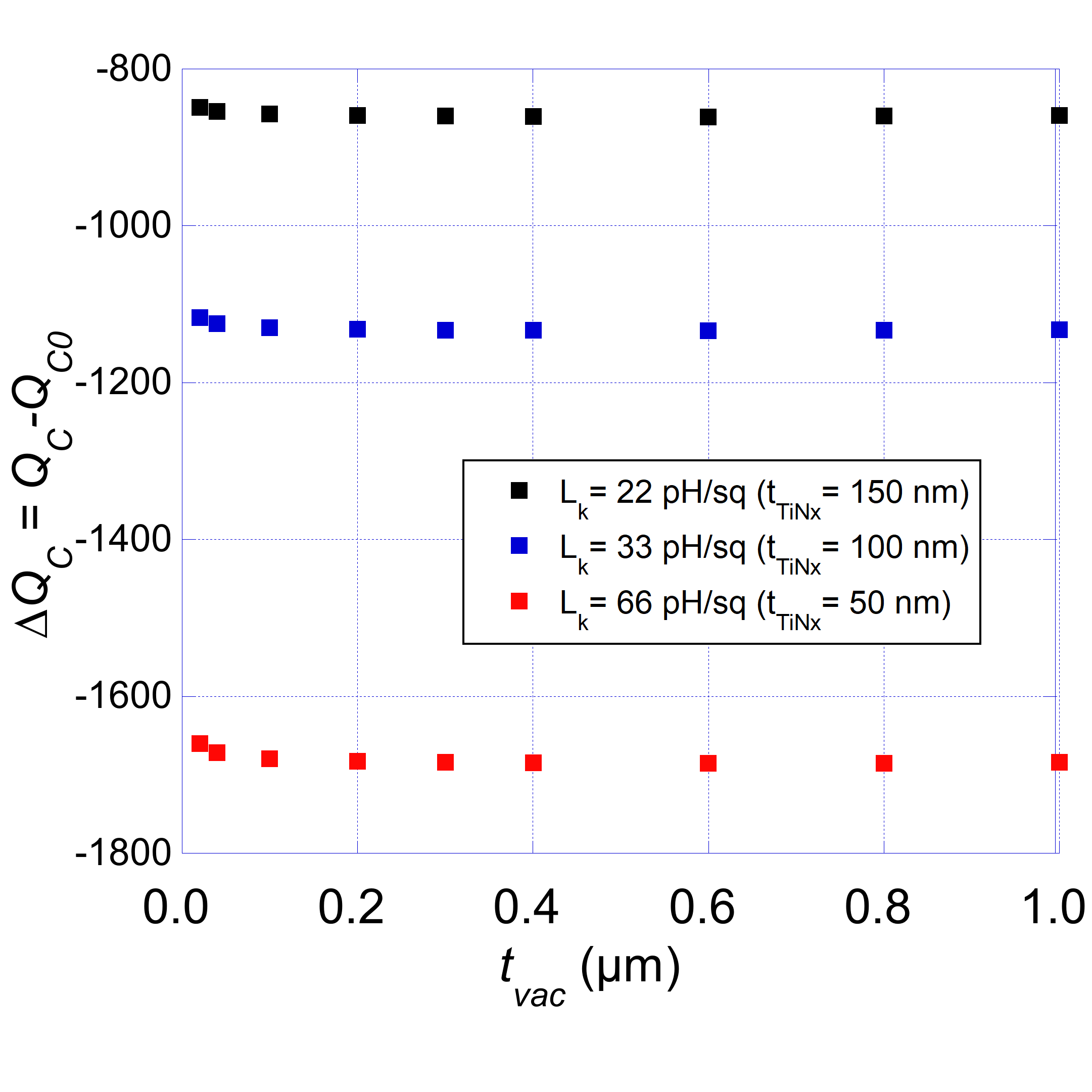}
\caption{}
\label{fig3:subim2}
\end{subfigure}
\caption{ (a) Resonance frequency $\Delta f_r$ and (b) coupling quality factor $\Delta Q_c$ shifts as a function of the vacuum gap and absorber thickness. Simulated with Nb/Au reflector, $f_{r0} \approx [4.13, 3.85, 3.47]$~GHz and $Q_{c0} \approx [8000, 10000, 13000]$ computed from the unsuspended configuration for respectively $L_k \approx [22, 33, 66]$~pH/$\square$.}
\label{fig:image3}
\end{figure}

The addition of reflectors made of gold (Au) or aluminum (Al) beneath the suspended meander would improve the optical coupling by reflecting back the transmitted optical photons \cite{Nicaise_2022}. The gap between the meander and the reflector must satisfy the quarter-wave condition for optimal optical coupling $t_{vac}=\lambda/4n$ with $n=1$. Thus, between
$\lambda =$ 400 and 1000~nm, the optimal vacuum gap varies between  $t_{vac} = 100$ and $250~\text{nm}$. 
This vacuum gap range is probably still achievable using the fabrication process implemented to build vacuum capacitor \cite{Boussaha_2020}. 
Finally, a Nb layer is added below the Al or Au reflective layer to minimize the electrical loss with the superconducting proximity effect. To maximize this effect, the thickness of the Nb layer $t_{Nb}$ must be no less than its coherence length $\zeta_{Nb} \approx 40~\text{nm}$ \cite{nicaiceP_2022}.
 
\section{Simulation Results}\label{sec3}

We simulate different KID configurations, with and without a suspended meander, to investigate the frequency operation. Each layer of the reflector is progressively added to study their effects on the resonant frequency $f_r$ and coupling quality factor $Q_c$.
This study stresses on $Q_c$ as the total quality factor $1/Q = 1/Q_i + 1/Q_c$. For $Q_c << Q_i$, the total quality factor $Q$ can be approximated as $Q \approx Qc$.
The unsuspended meander-based KID (taken as a reference) resonates at $f_{r0}= 3.85~\text{GHz}$ with a coupling quality factor of $Q_{c0} \approx 10000$. Once the meander is suspended with a vacuum gap of 200 nm, the resonant frequency shifts to higher frequencies with $\Delta f_r = f_r-f_{r0} \approx +168~\text{MHz}$ (Fig.~\ref{fig2:subim1}) and the coupling quality factor slightly decreases $\Delta Q_c = Q_c-Q_{c0} \approx -1000$. This is obviously due to the vacuum gap which reduces the overall parasitic capacitance $C_p \propto \epsilon_{eff}$ within the meander as well as the meander/substrate interface as $\epsilon_{r, vac}= 1 \leq \epsilon_{eff} \leq \epsilon_{r, Al_2O_3}$. Albeit modest, this will probably reduce TLS noise generated by these parasitic capacitances. By adding a single (Au or Al) or bilayer (Nb/Au or Nb/Al) reflector, the frequency shifts towards lower frequencies (Fig.\ref{fig2:subim1}) of only 0.2 MHz ($\Delta f_r \approx +168~\text{MHz}$). The coupling quality factor remains around $\Delta Q_c \approx -1000$.

Here, the vacuum gap between the reflector and the meander is also set at 200 nm. This shift in frequency highlights the characteristic behavior of a parasitic capacitance generated this time by the stack made of a reflector, vacuum, and meander which form a metal/vacuum/metal plate capacitor. However, as shown in Fig. \ref{fig2:subim1}, compared with the reference (i.e. the unsuspended meander), the resultant parasitic capacitors produce an overall shift towards higher frequencies indicating that the stack features a weak parasitic capacitor. Unlike reflector-based KIDs which use dielectrics \cite{Nicaise_2022}, this will probably not introduce additional TLS noises. Furthermore, the contribution of each layer is not differentiated regardless of the number of reflector layers, or metal type as the parasitic capacitance is made of vacuum with a lower dielectric constant $\epsilon_{r, vac} = 1$ $\leq$ $\epsilon_{r, \text{Al}_2\text{O}_3}= 9.9$ (Fig. \ref{fig2:subim1}).

\begin{figure}
    \centering
    \includegraphics[width=0.7
    \textwidth]{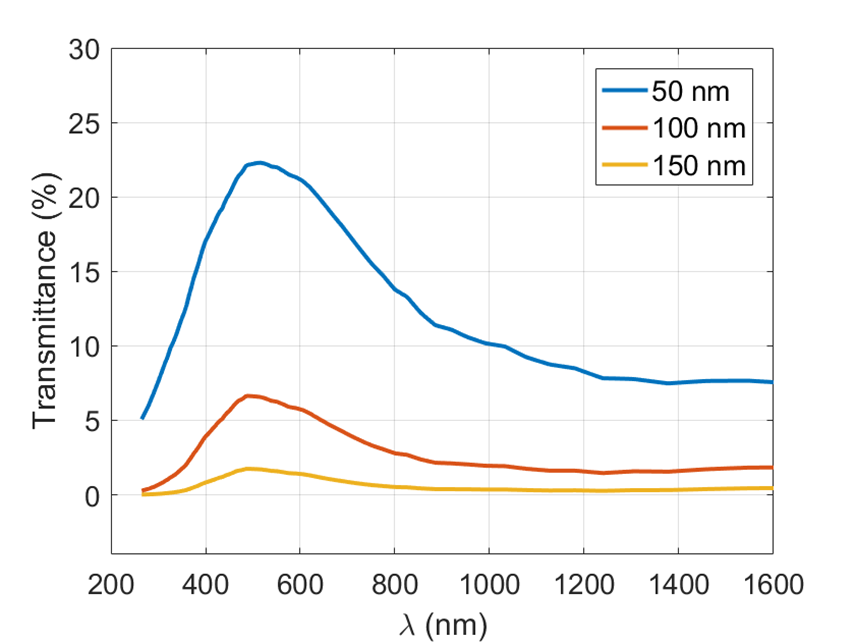}
    \caption{Simulated transmittance of a single TiN layer for three different
TiN thicknesses.}
    \label{fig:image4}
\end{figure}

Besides, a uniform current density is required to achieve a uniform phase response to incoming photons only within the meander. Fig. \ref{fig2:subim2} shows that it is well uniform in the suspended strips compared to the reflector. Furthermore, to assess the effect of meander suspension with a Nb/Au reflector, we carried out simulations for different vacuum gaps and meander thicknesses. As shown in Fig. \ref{fig3:subim1}, \ref{fig3:subim2}, as expected, for small gaps up to 200 nm where parasitic capacitors are more important, there is a stronger dependency of $\Delta f_r$ and, to a lesser extend, $\Delta Q_c$ on the vacuum gap, whereas, from 200 nm, they remain almost constant. These aspects will be taken into account especially when designing large-scale arrays with a resonance frequency step of 2-4 MHz between each resonance \cite{Jie_2023}. 

As reported in \cite{Szypryt2016}, only a fraction of the light is absorbed by the meander. The other fraction is either transmitted or reflected. To achieve optimum detector response, the kinetic inductance of the meander should be high and its volume should be small \cite{Boussaha_2023}. On top of this, the film absorption should ideally be total.  However, for thin TiN layers of a few tens of nm such as those shown in Fig. \ref{fig:image4}, transmittance becomes non-negligible. The suspended meander design would make it possible to recover the transmitted photons that have passed through the meander to be reflected back towards it thanks to a reflector placed at $\lambda/4$. The effectiveness of this concept will therefore depend considerably on below the transmission of TiN. Fig. \ref{fig:image4} shows the simulated transmittance, obtained by modeling the layer as a transmission line and retrieving the transmission parameter \cite{nicaiceP_2022}, for a single TiN layer for 50, 100 and 150 nm-thick. We notice that the transmittance decreases considerably with the TiN thickness. It decreases by around 70\% when the thickness increases from 50 to 100 nm and by around 90\% at 150 nm. This prompts the use of the thinnest TiN thickness which will be determined mainly by fabrication constraints.

\section{Conclusion and Further Work}\label{sec13}

By simulation, we have demonstrated the feasibility of an optical suspended meander-based KID design in order to improve response by trapping hot phonons. A superconducting reflector is also placed
under the suspended meander at a distance of a quarter wavelength ($\lambda/4$) to improve optical coupling. Compared to the unsuspended configuration, the suspended meander design affects both the resonance frequency and coupling quality factor. The fabrication of suspended meanders is obviously very challenging, especially for thin TiN layer, between 50 and 100 nm-thick, required to improve optical coupling using the reflective layer. This will be done using the process implemented for vacuum capacitor-based
KIDs \cite{Boussaha_2020}. In parallel, we will also focus on the modeling of the whole meander/vacuum/reflective
layer structure using the model of \cite{Nicaise_2022}.

\raggedbottom

\nocite{*} 
\bibliography{references}

\end{document}